# Security and Privacy Challenges in Cognitive Wireless Sensor Networks

**Jaydip Sen**
*Innovation Lab, Tata Consultancy Services Ltd., Kolkata, India*

## ABSTRACT

Wireless sensor networks (WSNs) have attracted a lot of interest in the research community due to their potential applicability in a wide range of real-world practical applications. However, due to the distributed nature and their deployments in critical applications without human interventions and sensitivity and criticality of data communicated, these networks are vulnerable to numerous security and privacy threats that can adversely affect their performance. These issues become even more critical in cognitive wireless sensor networks (CWSNs) in which the sensor nodes have the capabilities of changing their transmission and reception parameters according to the radio environment under which they operate in order to achieve reliable and efficient communication and optimum utilization of the network resources. This chapter presents a comprehensive discussion on the security and privacy issues in CWSNs by identifying various security threats in these networks and various defense mechanisms to counter these vulnerabilities. Various types of attacks on CWSNs are categorized under different classes based on their natures and targets, and corresponding to each attack class, appropriate security mechanisms are also discussed. Some critical research issues on security and privacy in CWSNs are also identified.

## INTRODUCTION

Over the last decade, wireless sensor networks (WSNs) have attracted a lot of interest in the research community due to their wide range of potential applications. A WSN consists of hundreds or even thousands of small devices each with sensing, processing, and communication capabilities to monitor a real-world environment. They are envisioned to play an important role in a wide variety of areas ranging from critical military surveillance applications to forest fire monitoring and building security monitoring (Akyildiz et al., 2002). Most of the WSN deployments operate in the unlicensed ISM bands (2.4GHz). Several other small range wireless protocols like Wi-Fi, Bluetooth etc. also use the same band. This has led to overcrowding in this band with the increasing deployment of WSN-based applications. As a result, coexistence issues in the ISM bands have attracted extensive research attention (Cavalcanti et al., 2007).

The increasing demand for spectrum in wireless communication has made efficient spectrum utilization a big challenge. To address this important requirement, *cognitive radio* (CR) has emerged as the key technology. A CR is an intelligent wireless communication system that is aware of its surrounding environment, and adapts its internal parameters to achieve reliable and efficient communication and optimum utilization of the resources (Mitola, 2000).With the advent of CR technology, we have a different perspective of the traditional WSNs. In the current cognitive wireless sensor networks (CWSNs), the nodes change their transmission and reception parameters according to the radio environment. Cognitive capabilities are based on four activities: (i) monitoring of spectrum sensing, (ii) analysis and characterization of the environment, (iii) optimization of the best communication strategy based on different constraints such as reliability, power, security and privacy issues etc., and (iv) adaptation and collaboration strategy. The cognitive technology will not only enable access to new spectrum but it will provide better propagation characteristics leading to reduction in power consumption, network life-time and reliability in a WSN. With cognitive capabilities, WSN will be capable of finding a free channel in the unlicensed band to transmit or could find a free channel in the licensed band for communication. A CWSN, therefore, will be able to provide access not only to new spectrum bands in addition to the available 2.4 GHz band, but also to the spectrum band that has better propagation characteristics. If a channel in a lower frequency band is accessed, it will certainly allow communications with higher transmission range in a CWSN, and hence fewer sensor nodes will be required to provide coverage in a specific area with a higher network life-time due to lower energy consumption in the nodes. CWNs will also provide better propagation characteristics by adaptively changing systems parameters like modulation schemes, transmit power, carrier frequency and constellation size. The result will be a more reliable communication with reduced power consumption, increased network life-time and higher reliability and enhanced *quality of service* (QoS) guarantee to applications.

In spite of the several advantages and benefits that CWSNs will bring forth (Cavalcanti et al., 2008), ensuring security will be a major challenge in these networks. Unless these challenges are solved to an effective level, deployment of CWSNs in real-world applications will face a serious impediment. As observed in (Burbank, 2008), the CR nature of a system introduces an entirely new gamut of threats and vulnerabilities that cannot be easily mitigated. These three salient characteristics of CR are its environmental awareness, learning and acting capabilities. Considering these characteristics from an attacker's perspective, a CWSN will provide much more capability to an attacker to launch attacks that are long-lasting and catastrophic in nature and those which can be triggered by simple spectral manipulations (Araujo, et al., 2012).

Security had already been an extensive area of research in WSN (Sen, 2009; Zhou et al., 2009; Martins & Guyennet, 2010). With the advent of CWSN and the perspective of security taking a much wider and complicated scope, it is obvious that research on the security aspects on CWSNs will attract even more attention of the research community. However, despite considerable amount of research on CR networks and the new interest in CWSNs, security in CWSNs has been a vastly unexplored area. Sensor data privacy will be another critical area which will be increasingly relevant as these networks find more applications in deployments that deal with sensitive and critical data.

Keeping this emerging trend of technology in mind, this chapter intends to provide a panoramic view of security and privacy-related issues in WSNs with particular focus on CWSNs. In the following sections, we present an extensive discussion on various security issues in WSNs and CWSNs and present their appropriate defense mechanisms based on the current state of the art.

## SECURITY AND PRIVACY ISSUES IN WSNS

In this section, we present a brief discussion on various security and privacy issues in a traditional WSN. All these issues are, however, applicable for CWSNs as well. We also discuss some defense mechanisms for handling these vulnerabilities.

Traditional WSNs are vulnerable to various types of attacks. These attacks can be broadly categorized into the following types (Shi & Perrig, 2004): (i) attacks on secrecy and authentication, (ii) attacks on network availability, and (iii) stealthy attacks on service integrity. Standard cryptographic mechanisms can prevent attacks on the secrecy and authenticity of the messages from outsider attacks such as eavesdropping, packet replay attacks, and modification or spoofing of packets. The attacks on network availability are also known as the *denial of service* (DoS) attacks. In stealthy attacks, an attacker compromises a sensor node and injects false data. DoS attacks, if launched successfully, can severely degrade the performance of WSNs. In the following, we describe how DoS attacks are made on different layers of the communication protocol stack.

**DoS attacks on the physical layer:** The physical layer is responsible for frequency selection, carrier frequency generation, signal detection, modulation, and data encryption (Akyildiz et al., 2002). Jamming in the physical layer is the most usual way to launch a DoS attack. In the jamming attack, the attacker interferes with the radio frequencies that the nodes in a WSN use for communication (Wood & Stankovic, 2002). The jamming attack is extremely catastrophic. Even with a less powerful jamming source, an adversary can potentially disrupt communication in an entire network by strategically distributing the sources of the jamming signal.

**DoS attacks in the link layer:** The link layer is responsible for multiplexing of data streams, data frame detection, medium access control, and error control. The attacks launched on this layer usually create collisions, resource exhaustion, and unfairness in allocation. A collision occurs when two nodes attempt to transmit simultaneously on the same frequency (Wood & Stankovic, 2002). An adversary may strategically cause collisions in specific packets such as ACK control messages. A possible result of such collisions is the costly exponential back-off. The adversary may simply violate the communication protocol and continuously transmit messages in an attempt to generate collisions.

**DoS attacks on the network layer:** The network layer of traditional WSNs is vulnerable to different types attack such as *spoofed routing information*, *selective packet forwarding*, *sinkhole*, *Sybil*, *wormhole*, *blackhole*, *grayhole*, *HELLO flood*, *Byzantine*, *information disclosure*, and *acknowledgment spoofing*. In the spoofed routing information attack, an attacker targets the routing information in the network by spoofing, altering or replaying the routing information to disrupt the traffic in the network. The disruptions include creation of routing loops, attracting or repelling the network traffic from selected nodes, extending or shortening the source routes, generating fake error messages, causing network partitioning, and increasing the end-to-end latency. In the selective forwarding attack, the attacker compromises a node in such a way that it selectively forwards some messages and drops the others (Wang et al., 2009a). In a sinkhole attack, an attacker makes a compromised node more attractive to its neighbors by forging routing information (Karlof & Wagner, 2003). The result is that the neighbor nodes choose the compromised nodes as the next-hop node to route their data through. This type of attack makes selective forwarding very simple as all traffic from a large area in the network flows through the compromised node. In the Sybil attack, a malicious node presents more than one identity in a network. This attack is particularly effective on routing algorithms, data aggregation, voting, fair resource allocation, and misbehavior detection. In the wormhole attack, a pair of malicious nodes first creates a wormhole. A wormhole is a low-latency link between two portions of a network over which one attacker node replays messages to the other attacker node (Karlof & Wagner, 2003). In the blackhole attack, a malicious node falsely advertises good paths (e.g., the shortest path or the most stable path) to the destination node during the path-finding process in reactive routing protocols, or in the route update messages in proactive routing protocols. A more delicate form of this attack is known as the grayhole attack, in which the malicious node intermittently drops data packets thereby making its detection more difficult. In an HELLO flood attack, an attacker uses a high-powered transmitter to fool a large number of nodes and makes them believe that they are within its neighborhood (Karlof & Wagner, 2003). Subsequently, the attacker node falsely broadcasts a shorter route to the base station and all the nodes that received the HELLO packets attempt to transmit to the attacker node. However, since these nodes are out of the radio range of the attacker, no communication will be established. In the Byzantine attack, multiple compromised nodes work in collusion and carries out attacks by creating routing loops, forwarding packets through suboptimal routes, and selectively dropping packets. In an information disclosure attack, a compromised node leaks confidential or important information to unauthorized nodes in the network. Such information may include data regarding the network topology, geographic location of nodes, or optimal routes to authorized nodes. In resource depletion attack, a malicious node attempts to deplete resources of other nodes in the network. The typical resources that are targeted are battery power, bandwidth, and computational power. The attacks could also be in the form of unnecessary requests for routes, very frequent generation of beacon packets, or forwarding of stale packets to other nodes. The acknowledgment spoofing attack is launched on routing algorithms that require transmission of acknowledgment packets. An attacking node may overhear packet transmissions from its neighboring nodes and spoof the acknowledgments, thereby providing false information to the nodes. Routing protocols in WSNs are also vulnerable to attacks such as touting table overflows, routing table poisoning, packet replication, route cache poisoning, and rushing attacks. A comprehensive discussion on these attacks can be found in (Sen, 2010a).

**DoS attacks on the transport layer:** The transport layer of WSNs is vulnerable to *flooding attack* and the *desynchronization attack* (Wood & Stankovic, 2002). If a protocol needs to maintain the state information at either end of an established connection, it becomes vulnerable to flooding attack. An attacker may repeatedly make new connection requests until the resources required by each connection are exhausted or reach a maximum limit. The desynchronization attack, on the other hand, attempts to disrupt an existing connection. For example, an attacker may repeatedly spoof messages to an end host causing the host to request retransmission of missed frames. The possible DoS and other types of attacks on WSNs and their corresponding countermeasures are listed in Table 1.

Table 1:  Various DoS attacks on the protocol stack of a WSN and their countermeasures

| Layer | Attacks | Defense Mechanisms |
|---|---|---|
| Physical | Jamming | Spread-spectrum, priority messages, lower duty cycle, region mapping, mode change |
| MAC | Collision | Error-correction code |
| | Exhaustion | Rate limitation |
| | Unfairness | Small frames |
| Network | Spoofed routing information & selective forwarding | Egress filtering, authentication, monitoring |
| | Sinkhole | Redundancy checking |
| | Sybil | Authentication, monitoring, redundancy |
| | Wormhole | Authentication, probing |
| | Hello Flood | Authentication, packet leashes by using geographic and temporal information |
| | Ack. Flooding | Authentication, bi-directional link authentication verification |
| Transport | SYN Flooding De-synchronization | Client puzzles, SSL-TLS authentication, EAP |
| Application | Logic errors Buffer overflow | Application authentication Trusted computing, Antivirus |
| Privacy | Traffic analysis, Attack on data privacy and location privacy | Homomorphic encryption, Onion routing, schemes based on traffic entropy computation, group signature based anonymity schemes, use of pseudonyms. |

**Node replication attack:** In this attack, the attacker attempts to add a node to a WSN by replicating (i.e. illegally copying) the node identifier of an already existing node in the network (Parno et al., 2005). A node replicated and joined in the network in this manner can potentially cause severe disruption in message communication in the WSN by corrupting the packets and forwarding them to wrong routes. This may also lead to network partitioning and communication of false sensor readings. In addition, if the attacker gains a physical access to the network, it is possible for him/her to copy the cryptographic keys and use these keys for message communication from the replicated node.

**Attacks on sensor data privacy:** Since in many applications WSNs are deployed for automatic data collection through efficient and strategic deployment of the sensor nodes, these networks are vulnerable to potential abuse of the collected data. Privacy preservation of sensitive data in WSNs is a particularly difficult challenge (Gruteser et al., 2003). Moreover, an adversary may gather seemingly innocuous data to derive sensitive information if he/she knows the aggregate data is collected from multiple sensor nodes. This is analogous to the "*panda hunter problem*", in which the hunter can accurately estimate the location of the panda by systematically monitoring the traffic (Ozturk et al., 2004). Some of the common attacks on sensor data privacy are: (i) eavesdropping and passive monitoring, (ii) traffic analysis, and (iii) camouflage. The most common form of attack on sensor data privacy is carried out by an attacker by silently listening to the messages communicated over the network. Eavesdropping on the control channels can be extremely effective strategy for an adversary to launch different attacks. The attacker may combine passive eavesdropping with an active traffic analysis. For example, by an effective traffic analysis, an adversary can identify some nodes that have special roles delegated to them. The attacker can then launch a directed attack to these nodes. In a camouflage attack, an adversary compromises a sensor node and later on uses the victim node to masquerade as a normal node.

## Security Mechanisms in Traditional WSNs

Numerous security mechanisms are proposed by researchers for defending against various attacks on WSNs. In the following, we provide a very brief discussion on some of these mechanisms. Interested readers may refer to (Sen, 2009) for a detailed discussion on these security protocols and algorithms.

**Applications of cryptographic mechanisms:** Since most of the security mechanisms for WSNs use cryptography, selecting the most appropriate cryptographic mechanism is a critical issue. The cryptographic algorithms and protocols must meet the constraints of the sensor nodes and should be evaluated by their code sizes, data sizes, processing time, and computational power requirements. It was a popular belief for long that the code size, processing time, and power requirements of the public key algorithms such as Diffie-Hellman key exchange protocol or RSA signatures are too heavy for WSN nodes. However, subsequent studies have shown that it is feasible to apply public key cryptography in WSNs by right selection of algorithms and associated parameters, optimization, and the use of low-power techniques (Wander et al., 2005). *Elliptic curve cryptography* (ECC) is particularly suitable for WSNs since it provides the same level of security as the RSA algorithm using a much smaller key size, thereby reducing the processing and communication overhead. In general, however, the private key operations in the public key cryptographic schemes are still expensive and most of the private key-related operations are assumed to be either carried out by the base stations or on some selected sensor nodes which have higher computational resources.

**Key management protocols:** Since the existence of a robust and efficient key management protocol is an essential pre-requirement for successful operation of a cryptographic mechanism, design of attack-resilient key management schemes that meet the resource constraints in such networks is a challenging task. The goal of key management is to establish keys among the nodes in a secure and reliable manner and to support node addition and revocation. Due to the high computational overhead of most of the public key cryptosystems, majority of the existing key management schemes for WSNs are based on symmetric key cryptography. A large number of key management protocols for WSNs exist in the literature. A detailed discussion on key management in WSNs can be found in (Sen, 2009).

**Defense mechanisms against the DoS attacks:** Since DoS attacks can be launched at different layers of the protocol stack, the defense mechanisms at different layers follow different approaches. In the physical layer, jamming attack can be defended by employing variations of spread-spectrum communications such as *frequency hopping* and *code spreading* (Wood & Stankovic, 2002). In *frequency-hopping spread spectrum* (FHSS), signals are transmitted by rapidly switching a carrier among many frequency channels using a pseudo-random sequence that is known to both the transmitter and the receiver. As a potential attacker would not be able to predict the frequency selection sequence, it will be impossible for him/her to jam the frequency being used at a given point of time. Another approach for handling jamming attacks in WSN is to tolerate the attacks by correctly identifying the jammed part of the network and effectively avoiding the nodes in the affected part by routing messages around it. In the link layer, *frame collision attacks* are handled by using error-correcting codes (Wood & Stankovic, 2002). The resource exhaustion attacks are prevented by *applying rate-limiting admission control mechanism* in the *medium access* *control* (MAC) layer so that the requests from nodes that intend to exhaust the energy-reserves of a node are rejected. Use of *time-division multiplexing* is another approach to defend against energy exhaustion attacks (Wood & Stankovic, 2002). Time-division multiplexing eliminates the need of arbitration for each frame and solves the indefinite postponement problem in a back-off algorithm. The adverse impact of caused by an attacker who intermittently launches link layer attacks can be mitigated by the use of small frames since it reduces the amount of time an attacker gets to capture the communication channel.

**Defense against attacks on the routing protocols:** Numerous mechanisms are proposed for defending against attacks on the network layer and on the routing protocols of WSNs. Since a detailed discussion of these schemes is beyond the scope of this chapter, we provide only a very brief discussion on some of the current and popular mechanisms. A detailed discussion can be found in (Sen, 2009).

A popular way to prevent spoofing and alteration of the routing packets is to append a *message authentication code* (MAC) to the routing packets. To defend against replayed information, counters or time-stamps are used in the messages (Perrig et al., 2002). *Selective forwarding* attack may be prevented using *multipath routing* (Karlof & Wagner, 2003). Hu et al. have proposed a mechanism called "*packet leashes*" for detecting and defending against wormhole attacks (Hu et al., 2003). As shown in Figure 1, in a wormhole attack, two or more malicious nodes collude together by establishing a tunnel using an efficient communication medium (i.e., a wired link or a high-speed wireless connection). During the route discovery phase, the route request messages are forwarded between the malicious nodes using the established tunnel. Therefore, the request message that reaches first at the destination node is the one that is forwarded by the malicious nodes. Consequently, the malicious nodes are added in the path from the source to the destination. Once the malicious nodes are included in the routing path, the malicious nodes drop the packets, resulting in complete or partial DoS attack. Sen et al. have presented a cooperative detection scheme that exploits the redundancy in routing information in an ad hoc network to build a robust detection framework for identifying malicious packet dropping nodes (Sen et al., 2007a). In (Sen et al., 2007b), a cooperative grayhole attack detection mechanism is proposed that utilizes a robust distributed collaborative algorithm among the nodes in an ad hoc network. Di Pietro et al. proposed a mechanism for securing group communications in WSNs (Di Pietro et al., 2003). The protocol is known as LKHW (Logical Key Hierarchy for Wireless sensor networks) and it is based on *directed diffusion-based multicast* mechanism. Lazos and Poovendran propose a similar tree-based key distribution scheme (Lazos & Poovendran, 2002).

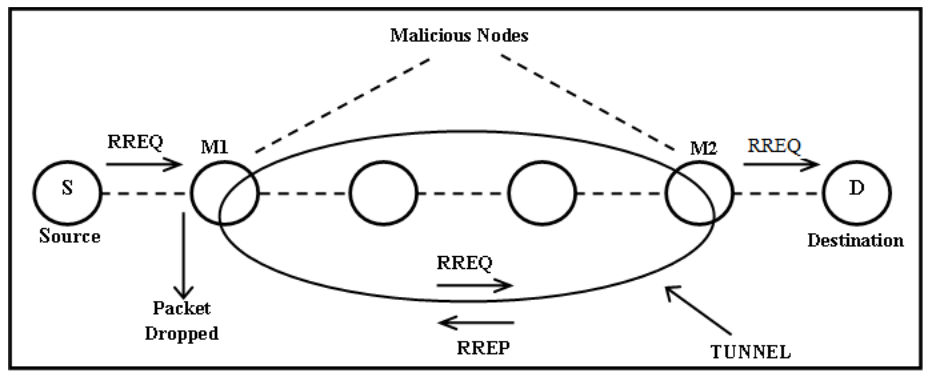

*Figure 1. Illustration of a wormhole attack launched by malicious nodes M1 and M2*

As discussed earlier in this section, most of the routing protocols for WSNs are vulnerable to various types of attacks such as: selective forwarding, sinkhole, blackhole, grayhole etc (Sen, 2010a). In the following, we briefly mention some of the well known propositions.

Liu & Ning propose a protocol called μTESLA– *micro version of the time, efficient, streaming, loss-tolerant authentication protocol*- for providing broadcast authentication in WSNs (Liu & Ning, 2004).

The protocol introduces an asymmetry through a delayed disclosure of the symmetric keys, leading to an efficient broadcast authentication scheme. Zhu et al. have proposed a scheme known as LEAP (Localized Encryption and Authentication Protocol) that is based on construction of a one-way key-chain for one-hop broadcast authentication (Zhu et al., 2003). In this scheme, each node generates a one-way key chain of certain length and transmits the first key of the key chain to each of its neighbor encrypting it with their pairwise shared keys. Deng et al. propose an "*intrusion-tolerant routing protocol in wireless sensor networks*" (INSENS) that adopts a routing-based approach to security in WSNs (Deng et al., 2002). A suite of security protocols called "SPINS" is proposed in (Perrig et al., 2002). Du et al. investigated the possible use of public key cryptography in designing secure routing protocols for WSNs (Du et al., 2005). Tanachaiwiwat et al. propose a secure routing protocol called "*trusted routing for location aware sensor networks*" (TRANS) that uses a symmetric key cryptographic scheme based on loose-time synchronization mechanism to ensure message confidentiality (Tanachaiwiwat et al., 2003). Papadimitratos et al. present a secure route discovery protocol that guarantees correct topology discovery in a WSN (Papadimitratos & Haas, 2002). The protocol relies on the use of MAC and an accumulation of the node identities along the route for a secure network topology discovery.

**Defense against attacks on the transport layer:** For defending against the flooding DoS attacks at the transport layer, Aura et al. proposed the use of "*client puzzles*" (Aura et al., 2001). In client puzzle-based schemes, each client has to demonstrate its commitment to the connection by solving a puzzle before it can access any resource in a server. Since an attacker does not have infinite resources, it will be impossible for him/her to create new connections fast enough to cause resource starvation on the serving node. A possible defense against desynchornization attacks is to enforce a mandatory requirement of authentication of all packets communicated between the nodes (Wood & Stankovic, 2002). If the authentication mechanism is secure, an attacker will be unable to inject any spoofed message.

**Defense against the Sybil attack:** A defense mechanism against the Sybil attack must ensure that a framework is in place that can validate a particular identity is only being held by a given physical node (Newsome et al., 2004). *Random key pre-distribution* techniques (Eschenauer & Gligor, 2002) can effectively be used to defend against the Sybil attack. In random key pre-distribution, a random set of keys or key-related information is assigned to each sensor node so that in the key setup phase, each node can discover or compute the common keys shared by it with its neighbors. The common keys are used as shared secret session keys to ensure node-to-node secrecy.

**Defense against node replication attack:** Parno et al. have proposed a mechanism for distributed detection of node replication attacks in WSNs (Parno et al., 2005) using two approaches: (i) *randomized multicast* and (ii) *line-selected multicast* – both of which are based on collaborative participation of multiple sensor nodes. The randomized multicast algorithm distributes location information of a node to randomly selected *witnesses* and exploits the *birthday paradox* to detect replicated nodes. The line-selected multicast algorithm is based on *rumor routing* (Braginsky & Estrin, 2002), and it uses network topology-related information to detect node replication.

**Defense against the traffic analysis attack:** Deng et al. propose a mechanism for defending against traffic analysis attacks in WSN (Deng et al., 2005a). The authors have identified two different classes of traffic analysis attacks: (1) *rate monitoring attack* and (2) *time correlation attack*. In rate monitoring attack, an adversary first monitors the packet sending rate of the nodes in its neighborhood, and then moves closer to the nodes that have a higher packet sending rate. In a time correlation attack, the adversary observes the correlation in sending times between a node and its neighbor node that is assumed to be forwarding the same packet and deduces the path by following each forwarding operation as the packet propagates towards the base station. Deng et al. have proposed mechanisms to defend against both these attacks (Deng et al., 2005a).

**Defense against attacks on sensor privacy:** Since protection of privacy of sensitive data in the sensor nodes in WSNs is an important requirement in many applications, several schemes for this purpose have been proposed by the researchers. These schemes can be broadly divided into three categories: (1) anonymity schemes, (2) policy-based schemes, and (3) schemes based on information flooding. An anonymity scheme depersonalizes the data before it is released from its source. Gruteser et al. present an analysis on the feasibility of anonymizing location information in location-based services in an automotive telematics environment (Gruteser & Grunwald, 2003). Beresford et al. propose various anonymity techniques for an indoor location system based on the *Active Bat* (Beresford & Stajano, 2003). Sen proposes an efficient and reliable routing protocol for wireless ad hoc and mesh networks that preserves user privacy while providing robust authentication for the users (Sen, 2010b). In policy-based defense mechanisms, decisions on access control and authentication are made on the basis of a specified set of privacy policies. Molnar et al. present the concept of private authentication and demonstrate its application in the *radio frequency identification* (RFID) domain (Molnar & Wagner, 2004). Hengartner et al. propose a framework of an access control on location information (Hengartner & Steenkiste, 2003).

Use of information flooding is a popular approach to achieve privacy in communication. Ozturk et al. propose modifications to WSN routing protocols for protecting the location information of a source node by using randomized data routing and a *phantom traffic generation* mechanism (Ozturk et al., 2004). Deng et al. address the problem of defending a base station against physical attacks by concealing the geographic location of the base station (Deng et al., 2005b). Xi et al. present an attack on the flooding-based phantom routing approach presented by Ozturk et al (Ozturk et al., 2004) and describe "*greedy random walk*" (GROW) protocol to reduce the chance of an eavesdropper successfully collecting the communicated location information (Xi et al., 2006). Li et al. propose a scheme that provides both content confidentiality and source-location privacy using a two-phase routing mechanism (Li & Ren, 2009).

**Secure data aggregation:** In a WSN, certain nodes - called the "*aggregators*" - are responsible for carrying out data aggregation operations so as to optimize the utilization of precious bandwidth of the wireless links. If an aggregator node or a sensor node is compromised, it is easy for an adversary to inject false data into the network. In absence of a robust authentication mechanism, an attacker can fool the aggregators into reporting false data to the base station. For securing the aggregation process in WSNs, two types of techniques are used: (1) plaintext-based protocols and (2) ciphertext-based protocols. The plaintext-based protocols operate on plaintext information. Hu et al. propose a secure aggregation protocol on plaintext data using μTESLA in which the sensor nodes are organized into a tree with the internal nodes acting as the aggregators (Hu & Evans, 2003). Chan et al. have presented a "*secure information aggregation*" (SIA) framework for sensor networks (Chan et al., 2007). Cam et al. propose an "*energy-efficient pattern-based data aggregation*" (ESPDA) protocol for WSNs (Cam et al., 2006). Du et al. propose a "*witness-based data aggregation*" (WDA) scheme for WSNs to ensure validation of data fusion nodes to the base station (Du et al., 2003). Secure aggregation of ciphertext data in WSNs is required to preserve the privacy of sensor nodes in many applications (Castelluccia et al., 2009; He et al., 2007). The propositions are based on a particular encryption transformation called "*privacy homomorphism*" (PH) that allows direct computations on encrypted data.

**Defense against physical attacks on nodes:** The sensor nodes in a WSN can be protected against possible tampering by tamper-proofing the physical packages of the sensors. Propositions are made by researchers for building tamper-resistant hardware in order to make the memory contents on the sensor chips inaccessible to a potential external attacker (Anderson & Kuhn, 1997). Deng et al. propose various approaches for protecting sensors by deploying components outside them (Deng et al., 2005b). In another work, Deng et al. discuss defense mechanisms against search-based physical attacks (Deng et al., 2002). Wang et al. present a modeling framework for "blind" physical attacks on WSNs (Wang et al., 2005). Seshadri et al. propose "*software-based attestation for embedded devices*" (SWATT) to detect a sudden change in the memory content of a sensor indicating the possibility of an attack (Seshadri et al., 2004).

**Trust management:** A popular approach to enforce a high-level of security in WSNs is to deploy trust- and reputation-based frameworks. Issues such as judging the quality and reliability of the sensor nodes and the wireless links, robustness of the data aggregation operation, correctness of the aggregator nodes, and timeliness in packet forwarding by the sensor nodes can be addressed very effectively with the help of trust-based systems. A comprehensive discussion on trust and reputation and various security mechanisms based on these concepts can be found in (Sen, 2010c). Here, we briefly mention some of these schemes. Pirzada et al. propose an approach for building trust relationships between the nodes in an ad hoc network based on their packet forwarding behavior (Pirzada & McDonald, 2004). Oram describes various methods of finding paths from a source node to a designated target node in a peer-to-peer computing paradigm (Oram, 2001). Extending this approach, Zhu et al. provide a practical approach for computing trust in wireless networks (Zhu et al., 2004a). Sen proposes a trust-based secure and efficient searching scheme for peer-to-peer networks that utilizes topology adaptation by the trusted nodes (Sen, 2011). Yan et al. discuss a trust-based security framework to ensure data protection and secure routing in an ad hoc network (Yan et al., 2003). Ren et al. present a probabilistic approach to model a distributed trust framework for a large-scale ad hoc network (Ren et al., 2004). Ganeriwal et al. propose a reputation-based framework for high-integrity sensor networks using the beta distribution for reputation representation, updates and integration (Ganeriwal & Srivastava, 2004). Liang et al. present models for robust aggregation algorithms which can be adapted for WSNs (Liang & Shi, 2008).

## SECURITY VULNERABILITIES IN CWSNS

In this section, we discuss various security vulnerabilities which are specific to CWSNs. In conventional WSNs, the transmission parameters can be changed and the radio frequency (RF) bands can be used in the limits which have been defined by pre-defined standards and spectrum regulations. The implementations in the hardware and firmware are done based on these specifications and they cannot be changed dynamically during network communications. A *cognitive wireless sensor network* (CWSN), other hand, can communicate in a wide range of spectrum bands and it is capable of changing its transmission parameters dynamically during network communication in response to the changes in the sensed radio spectrum environment, signals received from other sensor nodes in the network. This capability is gainfully utilized to realize innovative spectrum management approaches like *dynamic spectrum access* (DSA) in which the allocation of spectrum bands to communication services can change with time or space. An attacker can compromise a CWSN by breaking the DSA mechanism by implementing spectrum misuse or by exhibiting selfish behavior. For example, the attacker node can transmit in an unassigned band or it can ignore the cognitive messages sent by the other nodes in the CWSN. DSA can be implemented using various architectures for the CWSNs, each architecture having its own specific security vulnerabilities. In most of the centralized and distributed approaches of DSA, it is assumed that the participating nodes are altruistic and make logical decisions to optimize the use of the spectrum resources. However, such approaches make CWSNs vulnerable to security threats, where malicious cognitive nodes may exhibit selfish behavior. Hence identification of various possible attacks on CWSNs is critical in order to design appropriate security schemes to defend against those attacks. As an illustration of one security issue in CWSNs, Figure 2 shows a scenario of a *hidden node problem*, in which two cognitive nodes have different perceptions of the spectrum because of their different locations.

In addition to all the vulnerabilities in the traditional WSNs that we have discussed earlier in this chapter, CWSNs have many other security problems. Some of common attacks on CWSNs are: (i) attacks on the communication protocols, (ii) masquerading attacks, (iii) unauthorized access to the spectrum, (iv) physical attacks on the sensor nodes, (v) internal failures of the sensor nodes, (vi) power exhaustion attacks on the sensor nodes, (vii) attacks on the objective functions of the cognitive engine, (viii) attacks on the administrative policies of the sensor nodes, (ix) attacks on the cryptographic protocols and security schemes, and (x) attacks on the privacy of the sensor data. In the following, we briefly discuss the aforementioned threats and vulnerabilities in CWSNs.

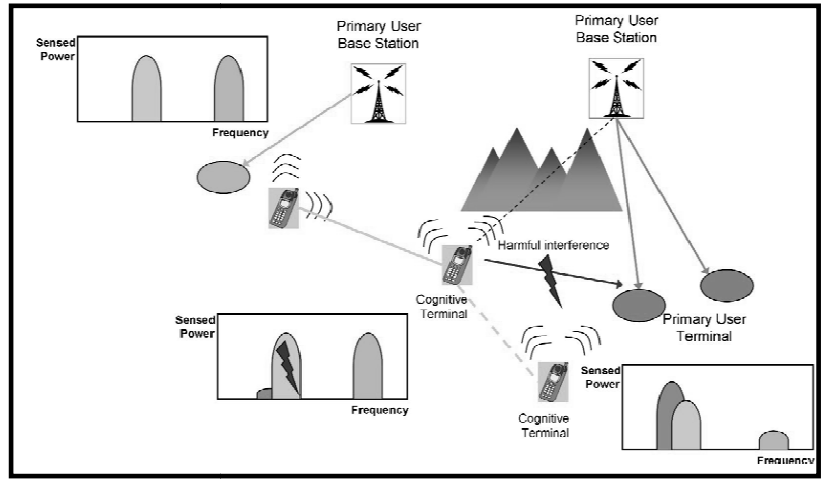

*Figure 2. An illustration of the hidden node problem in a CR network*

## Attacks on the Communication Protocols

The basic objective of the attacks under this category is to disrupt the communication in a CWSN. More specifically, the following attacks on the communication protocols in CWSNs can be identified: (i) replay attack, (ii) denial of service (DoS) attack, (iii) malicious alteration of the cognitive messages, (iv) Sybil attack, (v) hidden node problem, (vi) saturation of the cognitive control channels, (vii) eavesdropping of cognitive radio messages, (viii) disruption of the MAC, network layer, and cognitive engine of the cognitive radio network.

In a *replay attack* (Raymond et al., 2007), the attacker replays messages from earlier sessions of communications in the network. This attack is illustrated in Figure 3.The attacker may also send the replayed messages to another node which is not the intended recipient of the message. The receiver of the message, on finding that it is not the intended recipient forwards the messages further so that the message ultimately reaches the actual destination node. However, the delayed messages can lead to spreading of false information, since based on this delay, various characteristics of the network such as channel quality, network topology, routing etc, are computed. Since the nodes in a CWSN share extensive information among each other about various aspects of the network, spreading of false information can cause more damage in a CWSN than in a traditional WSN. For example, if the packets from the *primary users* (PUs) in a CWSN are replayed, the *secondary users* (SUs) might have a wrong perspective of the spectrum as well. This will forbid the SUs from using the frequencies and the protocols used by the attacker resulting in a sub-optimal and inefficient use of the network resources.

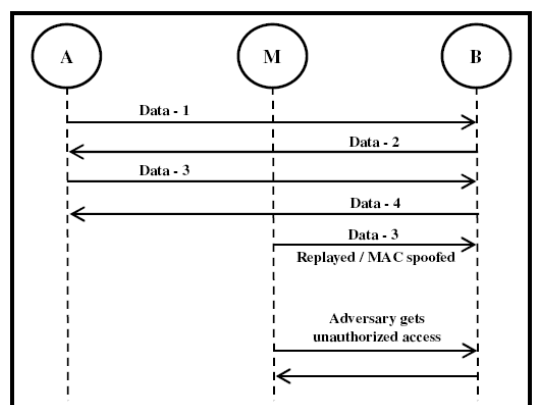

*Figure 3. Illustration of MAC spoofing and a replay attack launched by malicious node M*

In a DoS attack, the attacker makes the resources in the network unavailable to its legitimate users. There are many different ways in which a DoS attack may be launched – (i) jamming attack, (ii) collision attack, (iii) routing disruption attack, (iv) flooding attack etc.

In a *jamming attack*, the attacker transmits radio signals that interfere with the radio frequencies used by the nodes in a network. As discussed earlier, various ways of launching jamming attacks in WSNs and their defense mechanisms have been extensively studied by researchers over the last decade. In CWSNs, jamming attacks could be detrimental since it can rapidly exhaust the energies in the nodes and disrupt communication in the network. In a typical jamming attack in CWSN, a malicious node transmits signals at a high power using the PU frequency thereby disrupting the communication in the network. Jamming of the channels used to distribute cognitive messages in CWSNs is another serious threat. This attack can be launched against an out-of-bound *cognitive control channel* (CCC) or in-band CCC if the frequency of the channel is known. The objective of the *collision attack* is to violate the communication protocols used in CWSNs. While an attacker need not spend much energy in launching such an attack, the attack can cause serious damage in network services. Since the wireless medium is inherently broadcast in nature, detection of collision attacks and identification of the malicious nodes are non-trivial tasks. Since in CWSNs, the SUs share the spectrum, collision attack can easily and very effectively disrupt communications among the SUs. Hence, the collision attacks are more detrimental in CWSNs than in WSNs. In routing *disruption attack*, a malicious attacker does not forward the routing messages. The *grayhole* and the *blackhole* attacks are examples of these types of attacks. As already discussed in this chapter, these attacks are also possible in traditional WSNs. In a *flooding attack*, a malicious node sends a number of fake connection requests to a target victim node resulting in resource depletion in the latter. In *malicious alteration of cognitive message attack*, the adversary intentionally changes the cognitive messages in the network so that correct information cannot be exchanged among the nodes. The *Sybil attack* is launched by an attacker node that can assume multiple identities. This type of attacks can cause routing disruption, and unfair resource allocation in a resource sharing environment and in voting and reputation-based systems. For instance, the Sybil attack may be launched by a malicious node to generate additional reputations for malicious nodes or to change the information about the sensed spectrum. The *hidden node problem* arises when a CR node is in the protection region of an incumbent node but it fails to detect the existence of the incumbent. For example, if a CR node does not sense the presence of a primary user *base station* (BS) because of an obstacle, it transmits in the same frequency bands of the primary user, causing harmful interference. In the *saturation of the cognitive control channel attack*, the attacker launches a DoS attack against the *cognitive control channel* (CCC) by sending a large number of cognitive messages to the CCC thereby making the services of the control channel unavailable to other nodes in the CWSN. In the *eavesdropping of cognitive radio messages*, an attacker passively listens to the cognitive messages and subsequently uses the information contained on those messages to launch powerful attacks. In the attack involving *disruption of the MAC, network layer, and the cognitive engine of a CWSN*, an adversary attempts to target the protocols in the higher layers of the stack.

## Masquerading Attacks

This vulnerability involves the scenario in which a malicious adversary masquerades a primary user in a CWSN. The malicious attacker may mimic the primary user characteristics in a specific frequency band so that the legitimate secondary users erroneously identify the attacker as an incumbent and they avoid using that frequency band. This can be a selfish attack, because the attacker may subsequently use the frequency bands or launch a DoS attack to deny access to the spectrum resources to other secondary nodes in the CWSN. An example of the masquerading of a primary user in a CWSN is shown in Figure 4. A malicious CR node transmits a signal which is very similar to the primary user. On sensing this false signal, other CR nodes detect the presence of an additional primary user, and they avoid using the spectrum bands. In another form of a masquerading attack, a malicious CR node masquerades an honest node while collaborating with the other nodes in a CWSN to carry out important network functionalities such as: spectrum sensing, spectrum sharing, spectrum management, and handling of spectrum mobility.

This form of an attack can be dangerous since the malicious node may spread false information about spectrum sharing while participating in the collaborative decision making processes in a CWSN.

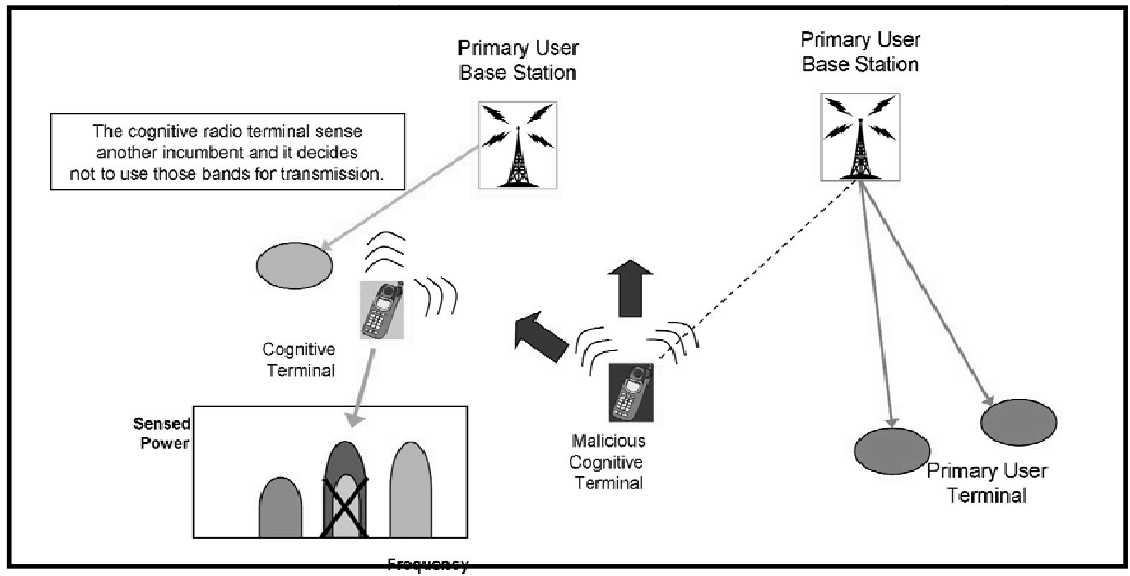

*Figure 4. An illustration of the masquerading of a primary user in a CR network*

Chen and Park introduced one important variation of masquerading attack that is known as the *primary user emulation* (PUE) attack (Chen & Park, 2006; Chen et al., 2008c). This attack is extremely effective in *dynamic spectrum access* (DSA) environments. In DSA, the primary users own licenses to different frequency bands and can use those bands whenever they wish. However, when the primary users are idle, the secondary devices can opportunistically use the spectrum on those bands. Such secondary users need spectrum sensing algorithms to detect when the primary user is active. To attack a DSA algorithm, an attacker needs to create a waveform that is sufficiently similar to that of the primary user so that a false positive may be triggered in the spectrum sensing algorithm. With the false signal being emitted by the attacker, the secondary users in the communication range of a primary user erroneously conclude that the primary user is active and vacate the channel. As a consequence, the adversary will gain unrivaled access to the specified frequency band. However, this attack is transient in nature since it is a *sensory manipulation attack* (Clancy & Goergen, 2008). Leon et al. have shown how a more sophisticated form of PUE attack can be launched when the attacker has some prior knowledge about the CR network (Leon et al., 2010). For example, if the attacker knows the exact time of occurrence of the *quite periods* of the CWSN, the attacker can launch the PUE attack during those time periods. Since during a quite period, all secondary users refrain from transmitting, if any user receives a signal during this period it assumes that the signal is emitted by a primary user. A malicious user may transmit during the quite period and effectively launch a PUE attack. Some DSA algorithms gather channel access information of the primary users and accordingly make prediction (Clancy & Walker, 2006). In such cases, spoofing primary user waveforms can lead to a *belief-manipulation attack* on the secondary users (Clancy & Goergen, 2008).

## Unauthorized Access of Spectrum

A malicious adversary node in a CWSN can launch an attack so that it can use spectrum bands for which it is not authorized or licensed and thereby gain more traffic capacity or bandwidth. Moreover, a malicious node can also emit power in unauthorized spectrum bands to cause DoS to the primary users.

## Physical Attacks on the Sensor Nodes

Physical attacks such as tampering with or damaging the hardware of even a small number of sensor nodes in a CWSN can prove to be catastrophic on the network operations. Since the functioning of a CWSN is dependent on the correctness of the critical information exchanged among the nodes, the adverse impact due to node compromise is more severe in CWSNs than in traditional WSNs. In a traditional WSN, compromise of a few nodes usually has a minor impact on the network performance since the connectivity is still maintained due to enough redundancy. However, successful operation of a CWSN is dependent on the distributed information among the nodes and their cooperative behavior. Hence, even a single compromised node in a CWSN can be a very powerful weapon for an attacker.

## Internal Failures of the Sensor Nodes

Failure of CR nodes in a CWSN may occur due to various reasons, i.e., memory fault, physical failure or other hardware failure. The impact of these failures may be quite damaging on the network services. For example, a malfunctioning CR node may transmit signals in a wrong frequency band or may not properly participate in important spectrum management-related collaborative decision making processes.

## Power Exhaustion Attacks on the Sensor Nodes

The sensor nodes are battery powered and energy constrained. To deplete the energies in the sensor nodes, an attacker can launch various types of power exhaustion attacks such as the *sleep deprivation attack* by engaging it in exchange of unnecessary message communications to quickly drain off its energy. If the attacker intelligently selects the target nodes, the failure of the nodes can cause network partitioning. In another form of power exhaustion attack, the attacker node can request a channel change very frequently causing a high rate of power usage in the target nodes.

## Attacks on the Objective Functions of the Cognitive Engine

In CWSNs, the cognitive engine in a sensor node has many radio parameters under its control. The cognitive engine determines the suitable values of these parameters over time in order to optimize its multi-goal objective functions (Clancy & Goergen, 2008). Various attacks are possible on the learning algorithms of the cognitive engines so that these algorithms produce suboptimal outputs. Since these attacks are targeted on the learning algorithms, they are also known as the *belief- manipulation attacks*. Clancy & Goergen have identified various input parameters of the cognitive engines such as: center frequency, bandwidth, transmit power, type of modulation, coding rate, channel access protocol, encryption algorithm, frame size etc (Clancy & Goergen, 2008). The cognitive radio may have three goals such as achieving *low-transmit power*, *high rate of transmission*, and *high security in communication*. Based on the application currently under use, the cognitive engine assigns different weights to these three goals to maximize its overall objective function. In order to build a robust framework, in the learning phase, the radio tries out various combinations of different values of the input parameters, measures the observed statistics of the network such as bit error rate, and then evaluates the objective function.

## Attacks on the Administrative Policies of the Sensor Nodes

The operating behavior of the sensor nodes in a CWSN are controlled by setting different policies in the nodes. These policies include security and privacy policies that determine access control, authentication, encryption/decryption, key revocation and other related operations. Several attacks can be launched on these policies such as: (i) *excuse attack* (ii) *newbie-picking attack* etc. (Araujo et al., 2012). If the security policies in the nodes are very generous and allow faster recovery of nodes that might have crashed or damaged and if these policies do not require nodes to prove their authenticity, a malicious node may exploit these policies by repeatedly claiming to have crashed/damaged thereby launching the *excuse attack*. In this way, wrong spectrum information can be sent to the network frequently to cause overload in the network. If a CWSN policy requires new nodes to pay their dues by making it mandatory for them to give information to the network for some period of time before they can consume any shared resource, a veteran node could move from one newbie node to another, leeching their information without being required to give any information back. This attack is known as *newbie-picking attack*.

## Attacks on the Cryptographic Protocols and Security Mechanisms

These attacks attempt to break the security mechanisms in the network and the nodes by compromising the cryptographic protocols used. The objectives of these attacks are to break the cryptographic algorithms, extract the keys used in encryption, decryption and hash computation, and to identify any possible vulnerability in the software and hardware of the nodes. Since the nodes in a CWSN are inherently resource constrained, the cryptographic schemes implemented in these nodes are light-weight in nature. These light-weight schemes sometimes prove inadequate against powerful and sophisticated attacks launched by high-end automated tools used by the attackers. An attacker may also launch attacks on the key management scheme in the network by using different strategies such as: naïve brute force attack, sophisticated dictionary attack, and passive session monitoring attack to capture important session-related information. One example of a sophisticated attack is the *differential power analysis* (DPA) attack, in which an attacker measures the strengths of the electromagnetic signals emitted from a target node to successfully identify the key used for encryption and decryption of messages.

## Attacks on the Privacy of the Sensor Data

Attacks on sensor data privacy are critical attacks since in many deployments of traditional WSNs and CWSNs, the sensor nodes collect and transmit sensitive data which need privacy protection. In CWSNs, the nodes share resources (i.e., spectrum) to establish communications among them and for developing a framework so that they are aware of the environmental parameters under which they are operate. If the privacy of such information is not protected, an adversary can successfully extract sensitive information from several nodes and may launch more powerful attacks on the network using the extracted information. The attacks on the privacy of the sensor nodes may involve different strategies such as: *eavesdropping*, *impersonation*, and *traffic analysis*. In passive eavesdropping attack, the attacker silently listens to the communications among the nodes to extract useful information about the session, and uses that information to launch a *replay attack* or an *impersonation attack*. In an impersonation attack, the attacker impersonates a legitimate node in the network and establishes communications with other nodes by providing its fake identity. An adversary node may monitor the messages to and from the legitimate nodes in a CWSN to launch a traffic analysis attack. The acquired information from traffic analysis is usually used later by the malicious adversary for devising more catastrophic attacks on the network. Location privacy threats represent a unique challenge in CWSN deployment. This is mainly due to the fact that a secondary user's spectrum sensing report on the signal propagation of the primary users are highly correlated to its physical location. Hence, similar to geo-locating individuals via WiFi or Bluetooth signals, a malicious attacker may exploit the correlation to geo-locate the secondary user and thus compromise the user's location privacy. Gao et al. have identified various types of attacks on the location privacy of the nodes in CR networks which are relevant in CWSNs as well (Gao et al., 2012).

## RELATED WORK ON IDENTIFICATION OF CWSN THREATS

In this section, we present a detailed discussion on the some of the existing works in the literature identifying various types of attacks on CWSNs and the mechanisms of launching these attacks.

**Jamming attack:** Sampath et al. present various ways in which jamming attacks can be launched on single channel and multi-channel 802.11 standard-compliant networks (Sampath et al., 2007). In the single channel jamming attack, the attacker continuously transmits high-power signals in the channel and causes interference to the communications from legitimate users in the network. In order to minimize energy consumption and to make the detection of the attack difficult, the attacker can also take a periodic jamming strategy in which the attacker transmits jamming packets at periodic intervals of time. In this strategy, the impact of jamming depends on the length of inter-jamming interval, the size of the jamming packets, and the size of the data packets sent to the victim node. It has been found that the impact of jamming degrades gracefully with the increase in inter-jamming interval, while the use of large packet size at the victim node increases the impact of jamming. In multi-channel jamming attacks, the attacker manipulates the CR to switch frequently across different channels and jam multiple channels simultaneously. Since, in addition to fast channel switching, the nodes in a CWSN have advanced channel sensing capabilities, the attacker can use a CR node to build up channel usage patterns of network users, and switch only among the channels are currently under use. These types of highly intelligent and efficient attacks are very difficult to detect in CR networks. Burbank et al. present a detailed description on how various types of jamming attacks can be targeted in a CR network and how adverse these attacks can be on the overall network performance (Burbank et al., 2008). All these attacks are relevant in CWSNs as well. Sethi and Brown discuss various ways in which DoS attacks can be launched on CR networks and present a framework to analyze those attacks (Sethi & Brown, 2008). The framework, known as the "*Hammer Model Framework*", graphically presents the potential risks sequences for DoS attacks, and investigates various types of vulnerabilities that may prevent CR communication is specific spectrum bands or completely deny a CR network to communicate or induce it to cause harmful interference to its existing legitimate users. In addition to jamming attacks, the authors have also considered attacks related to malicious alterations of cognitive messages and masquerading of a CR node by a malicious adversary. Zhang et al. propose a classification of various attacks on a CR network which can adversely affect its learning capability and its ability to gainfully utilize the benefits of dynamic spectrum access (Zhang et al., 2008). Arkoulis et al. have identified, analyzed and explained the security weaknesses and vulnerabilities of cooperative, dynamic and open spectrum access environments that can be targeted by a malicious adversary to disrupt the network operations or degrade its performance (Arkoulis et al., 2008). The authors have followed an approach for identifying threats based on the types of anomalous behavior of the nodes such as: misbehavior, selfishness, cheating, and malicious intention. Burbank presents some major security threats in CR networks in general, and identifies various challenges in defending against these threats (Burbank, 2008). In order to identify specific security challenges in CR networks which are applicable for CWSNs as well, the author has first pointed out two fundamental differences between a traditional wireless network and a CR network. In the CR networks the attacker has: (i) the potential far reach and long-lasting nature of an attack, and (ii) the ability to have a profound effect on network performance and behavior through simple spectral manipulation by generating false signals. In a CR network, the nodes exchange locally-collected information to construct a perceived environment that that determines the current and future behavior of the nodes. The author argues that in a CR network, a malicious adversary can propagate its behavior through the network in the same way a malicious worm propagates in a network. The adversary can carry out spectral manipulation for influencing the behavior of a set of local CRs or a distant CR as well. The author has also identified various features of CR networks and the implications of these features on potential attacks on these networks. Brown and Sethi present a multidimensional analysis and assessment of various DoS attacks on all types of CR networks (Brown & Sethi, 2007). The authors have carried out vulnerability analysis of CR network against various DoS attacks using different parameters such as network architecture employed, the spectrum access technique used, and the spectrum awareness model. The attacks are categorized into two types: denial attacks and induce attacks. While the denial attacks are intended to prevent communications in the network, the induce class of vulnerabilities stimulate the CR node to communicate causing interference with a licensed transmitter. The adverse impact of these attacks is not reflected immediately. However, these attacks cause permission policies to be tightened or eliminated potentially denying network services over a long-term. In multi-dimensional analysis of DoS attacks, a number of metrics such as jamming gain, jamming efficiency, packet send ratio, and packet delivery radio, have been proposed.

**Primary user emulation (PUE) attack:** Chen et al. have identified a threat to spectrum sensing, named the *primary user emulation* (PUE) attack in which an adversary's CR transmits signals whose characteristics emulate those of incumbent signals (Chen et al., 2008c). This attack is particularly easy to launch in CWSNs due to the highly flexible and software-based air interfaces of CR sensor nodes. The PUE attack can be catastrophic since it severely interferes with the spectrum sensing process and reduces the channel resources available to the legitimate unlicensed users in the network. In another work, Chen et al. have discussed two different security threats on CR network which are known as *incumbent emulation* (IE) attack and *spectrum sensing data falsification* (SSDF) attack (Chen et al, 2008b). The IE attack is essentially same as the PUE attack since the primary users are also sometimes referred to as the incumbents. The SSDF attack is carried out by malicious secondary nodes that transmit false spectrum sensing data to other nodes. This attack is critical in a CWSN, since sending of false spectrum sensing information to a data collector in the network can cause the data collector to make a wrong spectrum sensing decision resulting in a catastrophic impact on the network performance.

Wang et al. have argued that one of the major challenges in CR networks is to detect the presence of primary users' transmission, since malicious secondary users can send false spectrum sensing information and mislead the spectrum sensing data fusion process to cause collision, interference and inefficient spectrum usage (Wang et al., 2009c). Clancy and Khawar have highlighted the need of robust signal classification mechanisms for CR networks so that different types of transmitters can be differentiated in a particular frequency band in order to defend against the PUE attacks (Clancy & Khawar, 2009). Anand et al. present a novel analytical framework to analyze the feasibility of PUE attack in a CR network which can be applied to a CWSN as well (Anand et al., 2008).

**Masquerading attack:** Masquerading attacks on a CR node and attacks involving malicious alteration of CR nodes for disrupting spectrum sensing functions have also been studied extensively. Wang et al. have shown the adverse effect of malicious and compromised secondary users in a CR network (Wang et al., 2009b). Chen et al. have considered the security issues related to malicious secondary users reporting false spectrum sensing information due to *Byzantine failure* in a distributed spectrum sensing environment in a CR network (Chen et al., 2008a). The Byzantine failures, such as device malfunction or attacks severely affect the spectrum sensing functions in a CR network since these failures or attacks can enable an attacker to constantly report the spectrum in a band being in use causing severe under-utilization of the available spectrum. Hu et al. have also addressed the issue of Byzantine failures of secondary users in a CR network, and have proposed a security mechanism that is similar WSPRT, in which the binary local reports used in WSPRT are replaced with *N*-bit local reports to achieve an enhanced detection performance (Hu et al., 2009). Mody et al. have discussed various security threats in IEEE 802.22 standard-compliant devices which are deployed in CR networks (Mody et al., 2009).

**Attacks involving false spectrum reports sent secondary users:** The threats due to false spectrum reports sent by malicious secondary users are also critical. Securing the control channels ensures that CR messages communicated over these channels cannot be altered by a malicious adversary. This protection is critical in CWSNs which are deployed for mission-critical applications. In this perspective, Safdar and O'Neill have identified the need of securing the cognitive control channels to perform channel negotiations before data communication among the nodes in a CR network (Safdar & O'Neill, 2009). The authors propose a novel framework for providing common control channel security for co-operatively communicating CR nodes so that a pair of CR nodes can authenticate each other. Li and Han have discussed a critical security issue in collaborative spectrum sensing in which malicious secondary user(s) sends false spectrum report to thwart the spectrum data fusion process (Li & Han, 2010).

**Attacks on cognitive control channels:** Securing cognitive control channels is an extremely important security issue in CR networks. Prasad have argued that design of a CR network poses many new technical challenges in protocol design, power efficiency, spectrum management, spectrum detection, environment awareness, novel distributed algorithms design for decision making, distributed spectrum measurements, *quality of service* (QoS) guarantees, and security (Prasad, 2008). The author have identified various research challenges for security in CR networks and have presented the security and privacy requirements, threat analysis and an integrated framework for security using fast authentication and authorization architecture. The particular focus of the proposed security framework is to defend against jamming attacks and attacks on the *cognitive control channels* (CCCs) in CR networks.

**Attacks on the MAC layer:** Zhu and Zhou have provided a security analysis of the MAC protocols used in CR networks by investigating the impact of DoS attacks on these protocols (Zhu & Zhou, 2008). In order to make a security analysis of the MAC protocols, the authors have distinguished two types of attacks and then discussed how DoS attacks can be successfully launched on the MAC protocols. The authors have also presented a detailed discussion on MAC layer greedy behaviors in CR networks and the factors that determines the efficiency of the DoS attacks.

**Attacks on the cognitive engines:** Clancy and Goergen identify three classes of attacks on the cognitive engine of CR networks (Clancy & Goergen, 2008). All these types of attacks manipulate the behaviour of the CR system such that the radio acts either sub-optimally or even sometimes maliciously. Three classes of attacks are identified: (i) sensory manipulation attacks against policy radios, (ii) belief manipulation attacks against the learning radios, and (iii) self-propagating behaviour leading to cognitive radio viruses. In a policy radio, the main vulnerability lies in the fact that an attacker can spoof faulty sensor information that can cause the radio to select a sub-optimal configuration. Since the radio sensors take digitized RF and extract useful statistics from it, by manipulating the RF that is available to the radio, an attacker can cause faulty statistics to appear in the CR knowledge base. The learning radios are also vulnerable to the same threats as the policy radios. However, since a leaning radio uses all its past experiences in building its long-term behvaior, attacks on it are much more detrimental. For example, an attacker can transmit a jamming signal whenever a policy radio attempts to switch to a faster modulation rate. This will always force the CR to operate at a lower modulation rate, resulting in lower links speeds and link degradation. The authors have called these attacks as *belief manipulation attack* since these attacks can potentially have long-term adverse impact of the learning radios. The self-propagating behaviour of the radio can be utilized by a malicious attacker to launch the most powerful type of attack. In such an attack, the state on radio causes a behaviour that can induce the same state on another radio. Once the target radio attains the state, it exhibits behaviour that leads to a state change in another radio so that it attains the same state. Eventually, the same state propagates through all radios in a particular area in the CR network. The net effect is that of a cognitive radio virus that propagates through the network.

**Threats related to the hidden node problem:** The threats related to the hidden node problem in CR networks have also been studied extensively by the researchers. Biswas et al. propose a technique to handle both wideband and cooperative spectrum sensing tasks in a distributed spectrum sensing environment (Biswas et al., 2009). Nuallain presents a fast and robust propagation method for addressing the hidden node problem in a CR network (Nuallain, 2008). Bliss have investigated the optimal spectral efficiency for a given message size that minimizes the probability of causing disruptive interference for a CR network (Bliss, 2010). The goal of the work is to have an optimization between longer transmit duration and wider bandwidth versus higher transmit power so as to tackle the hidden node problem.

It may be noted that among all the threats in CWSNs, jamming and masquerading of the primary users are most critical. Another interesting point to note is that some of the attacks can be correlated to launch a powerful *two-phase attack*. For example, an attacker may first eavesdrop on cognitive messages and then may replicate and modify the cognitive messages to transmit false information.

## SECURITY MECHANISMS TO DEFEND AGAINST ATTACKS IN CWSNS

In this section, we first identify the main security requirements in a CWSN and then discuss various security schemes for defending against various attacks. In a CWSN, the sensor nodes participate in collaborative spectrum sensing activities. Gao et al. have identified the following security requirements in CWSNs (Gao et al, 2012): (i) authentication mechanisms, (ii) incentive mechanisms, (iii) data and message confidentiality, (iv) privacy protection of the sensor data.

**Authentication mechanisms:** A robust authentication mechanism is a prime requirement in collaborative spectrum sensing. The authentication scheme may have different perspectives to different categories of nodes in a CWSN. The authentication of the primary users is a critical issue since an attacker may transmit signals with high power that has close resemblance with the signals of a primary user and launch a *primary user emulation* (PUE) attack (Chen et al., 2008c; Liu et al., 2010). To prevent such an attack, the secondary users should have a robust verification scheme for verifying the authenticity of the received signals. Similarly, when the secondary users receive the sensing reports from other users, they should be able to verify the authenticity of the other secondary users; otherwise, a potential adversary may be able to spoof the identity of a secondary user. The authentication of sensing reports distributed across the network is also a very important issue. Even if the authentications of the secondary users are done during the sensing report aggregation process, it is still possible for a malicious secondary user to send false sensing reports and launch *spectrum sensing data falsification* (SSDF) attack (Wang et al., 2009b). Hence, each sensing report in the aggregation process should be authenticated.

**Incentive mechanisms**: Most of the current collaborative sensing schemes assume that the secondary nodes voluntarily participate in spectrum sensing. However, this assumption may not hold good for selfish secondary users who may not cooperate in order to conserve their own resources (Wang et al., 2010). Such selfish behavior may seriously degrade the performance of a CWSN. Incentive schemes are necessary for minimizing the probability of such selfish behavior.

**Data and message confidentiality:** The sensing reports need to be well protected so that these messages are not misused by unauthorized external users. Data and message confidentiality can be achieved by using end-to-end robust encryption algorithms which in turn needs mutual authentication and authorization among the collaborating nodes participating in spectrum sensing.

**Privacy preservation of sensor data:** Privacy protection is primarily for preserving the anonymity of the sensing nodes and/or privacy of its location. Location privacy protection attempts to prevent a possible adversary form linking a sensing node's sensing report to the physical location of the sensing node.

In order to satisfy the aforementioned security requirements and to defend against various possible attacks on the sensor nodes in a CWSN, various defense mechanisms are proposed by the researchers. These security schemes can be broadly divided into the following categories: (i) mechanisms for enhancing the robustness in sensor inputs, (ii) schemes based on the reputation and trust of the nodes, (iii) mechanisms based on identification of masquerading attack by signal analysis, (iv) robust authentication schemes using appropriate cryptographic algorithms, (v) mechanisms for preventing unauthorized access to the spectrum, (vi) mechanisms for defending against attacks on the MAC layer and the cognitive engine of the network, (vii) schemes for increasing the robustness of the cognitive control channel against jamming and saturation attacks, and (viii) schemes using geo-location database of the primary users in the network. In the following, we present a brief discussion on these various types of security mechanisms.

## Enhancing the Robustness in Sensor Inputs

Many of the attacks on CR networks can be defended if the reliability of sensor inputs is enhanced. For example, if the cognitive radios can minutely identify the differences between interference and noise, they can distinguish natural and artificial RF events. Such sensors can feed specialized policy algorithms that specifically look for hostile signals that may be try to subvert a radio's belief. In a distributed computing scenario, a group of cognitive nodes can fuse sensor data to improve the performance of the overall network. For example, if multiple sensor nodes exchange time-synchronized RF information, they can cross-correlate the exchanged information to arrive at a more precise identification of an attacker. The task becomes challenging, however, since the all sensory inputs are imprecise to a certain extent.

## Reputation and Trust-based Security Systems

A significant number of schemes have been proposed by the researchers using reputation and trust of the CR nodes for defending various types of attacks. Using the concepts of reputation and trust, a CR node can be mapped to a particular level of reputation and trust on the basis of the spectrum sensing information the node shares with other CR nodes. If the information shared by the node is found to be not correct after a certain number of iterations, then the specific CR node is considered to be malicious and appropriate action is taken against the node based on pre-defined security policies.

Zeng et al. have proposed a reputation-based *cooperative spectrum sensing* (CSS) framework using trusted nodes in a CR network for achieving correctness in the global decisions on spectrum sensing (Zeng et al., 2010). In the proposed scheme, at the beginning, sensing information from trusted nodes is only considered reliable and used in the decision making. Reputations of other CR nodes are put in the pending state, and they are accumulated through a consistency check between the global and local sensing decisions. The information received from the nodes which have their trust values greater than a pre-defined threshold is then considered reliable and their sensing results are incorporated in the CSS. The use of reputation system increases the robustness of cooperative sensing scheme. Duan et al. propose a spectrum sensing algorithm that is based on the reputation of the nodes (Duan et al., 2009). The algorithm is effective in mitigating the effects of shadowing and fading in wireless channels and in eliminating the problem related to fail sensing in CR networks with double threshold detector. Li and Han have presented an anomaly detection algorithm for identifying attackers in a collaborative spectrum sensing environment (Li & Han, 2010). The proposed scheme does not assume any a priori information about the strategy used by the attackers in launching the attack, which makes scheme suitable for real-world deployment. Kaligineedi et al. propose an attack detection scheme to identify malicious users that send false spectrum sensing information in a CR network (Kaligineedi et al., 2008). The proposed scheme uses the average power obtained from the real-valued reports received from the CR nodes for making a global decision on spectrum sensing. Chen et al. present a security scheme based on *weighted sequential probability ratio test* (WSPRT) to address Byzantine failures on CR nodes in the data fusion process of collaborative spectrum sensing (Chen et al., 2008a). The mechanism involves an allocation of a reputation rating to each node based on the consistency of its local sensing report with the final decision in the spectrum sensing. Peng et al. discuss various security aspects in cross-layer design of CR networks and propose a novel architecture in which dynamic channel access is achieved by a cross-layer design between the PHY and the MAC layers of a CR network (Peng et al., 2009). The proposed architecture is able to handle Byzantine failure of nodes. Anand et al. have analyzed the performance limitations of collaborative spectrum sensing in a DSS environment under Byzantine attacks where malicious users send false spectrum sensing data to the fusion center leading to increased probability of incorrect sensing results and wrong global decisions being taken by the CR (Anand et al., 2010). It has been shown that if the percentage of Byzantine attackers in a CR network exceeds a certain threshold value, the data fusion scheme become completely incapable in carrying out reliable data fusion and no reputation-based fusion system can achieve any performance gain in the data fusion operation. The authors have presented optimal attacking strategies for a given set of attacking resources and have also proposed possible counter measures at the data fusion center. Xu et al. propose a collaborative sensing algorithm that uses an energy detector with double thresholds and an extended data fusion rules to identify untrusted and possibly malicious CR nodes (Xu et al., 2009). Yu et al. have studied the security issues related to the SSDF attack in which attacker(s) sends false local spectrum sensing results in a DSS environment (Yu et al., 2009). A consensus-based cooperative spectrum sensing scheme is proposed that is inspired from the self-organizing behavior of animal groups.

## Detection of Masquerading Attacks by Signal Analysis

Signal analysis is an important technique used in identification of malicious attacker(s) in CR networks. This method is very effective in addressing security threats which involve a malicious attacker masquerading as an incumbent transmitter by transmitting unrecognized signals in one of the licensed bands and thereby effectively preventing secondary users in the CR network from accessing the same spectrum band. Spectrum sensing can be done in a variety of ways. Some of the commonly used spectrum sensing methods are: energy detector based sensing (also known as radiometry or periodogram), waveform-based sensing, cyclostationarity-based sensing, radio identification-based sensing, matched filtering, multi-taper spectral estimation, wavelet transform-based estimation, Hough transform, and time-frequency analysis (Yucek & Arslan, 2009). However, each of these spectrum sensing techniques have vulnerabilities in a CR network since an adversary can masquerade a primary or a secondary user or by emulating its signal. Various security schemes have been proposed by researchers to detect and defend against such attacks. Some of the mechanisms are briefly discussed in the following.

Chen and Park propose a security mechanism for defending against masquerading of a primary user by a malicious adversary (Chen & Park, 2006). The proposed scheme is based on a transmitter verification procedure that employs a location verification scheme to distinguish incumbent signals (i.e., signals from a primary user) from unlicensed signals masquerading as incumbent signals. Location verification is achieved by using two techniques: (i) *distance ratio test* (DRT), which uses the *received signal strength indicator* (RSSI) of a signal source and (ii) *distance difference test* (DDT), which uses relative phase difference of the received signal as the signal is received at different receivers. It is assumed that the location information of some of the CR nodes in the network is always known a priori either because these nodes are fixed or they use trusted GPS information. These CR nodes perform DRT and DDT operations within their coverage areas and also serve as the *location verifiers* (LVs). The LVs exchange the location information of incumbent transmitters through a cognitive pilot channel. Zhao and Zhao propose a cooperative detection scheme that can suppress malicious users (Zhao & Zhao, 2009). In the proposed scheme, the secondary users collaborate by exchanging and using decision fusion on the local decision results instead of using the detected energy. A mechanism of weighted coefficients is used which updates the weights of the coefficients recursively according to the deviations between separate decision information and the combined final results. Zhao et al. propose an identification mechanism of the CR nodes using an analysis of the transmitted signals in which *wavelet transform* is used to magnify the fingerprints of the transmitter characteristics (Zhao et al., 2010). This PHY-layer authentication approach is intended to prevent the PUE attack in CR networks. Afolabi et al. have describe a PHY layer attack model that exploits the adaptability and flexibility of the CR networks and propose a *waveform pattern recognition scheme* to identify emitters and detect camouflaging attackers by using *electromagnetic signature* (EMS) of the transceiver (Afolabi et al., 2009). The EMS of a device is computed based on the distinctive behavior in the waveform being emitted by the components of the transceiver including the frequency synthesis systems, modulator sub-systems, and the RF amplifiers. Clancy and Khawar present sophisticated signals processing algorithms like cyclostationary analysis, classification engines, or signal feature extraction for identifying false signals in CR networks (Clancy & Khawar, 2009). The authors propose the use of unsupervised learning in feature-based signal classification and provide recommendations to mitigate the impact of the attack on CR networks.

## Robust Authentication using Cryptographic Techniques

Cryptographic techniques are widely used in designing robust and efficient authentication protocols in wireless networks. However, in CWSNs, authentication mechanism should be adaptable to all communication protocols with which the CR nodes have to interface. Hence, design and implementation of authentication protocols for CWNs pose significant challenges.

Kuroda et al. propose a radio-independent authentication framework for CR networks that can be integrated with the *extensible authentication protocol* (EAP) (Kuroda et al., 2007). The protocol is suitable for deployment in real-world networks since it allows fast switchover in CR network and does not need any communication with the *authentication authorization and accounting* (AAA) server for any re-authentication of the CR nodes. Jakimoski and Subbalakshmi have proposed an efficient and provably secure protocol that can be used to protect the spectrum decision process against a malicious adversary (Jakimoski & Subbalakshmi, 2009). The proposed protocol is designed to provide secure spectrum decisions in a clustered infrastructure-based network where the spectrum decisions are made at periodic intervals and the decision in each cluster is taken independently of the decisions in other clusters.

The CWSNs should ensure authorization of the cognitive sensor nodes for transmitting specific spectrum bands or for performing specific network functions. The authorization is often conditional to the nature of the spectrum environment, i.e., the presence of primary users in the area. The authorization is needed to define the roles of the CR nodes in performing the CR functions. For authentication and authorization purposes, the nodes exchange information through a common CCC. Safdar and O'Neill propose a security framework for protecting the information exchanged over the CCC (Safdar & O'Neill, 2009).

## Security Mechanisms for Prevention of Unauthorized Spectrum Access

A malicious node can access spectrum in a CR network in an unauthorized way either to use the spectrum selfishly or to launch a DoS attack on the primary users. Several propositions are made by researchers for defending against such attacks. In the following, we provide a brief discussion on some of these schemes. Xu et al. present a framework known as TRIESTE (Trusted Radio Infrastructure for Enforcing SpecTrum Etiquettes) for ensuring that radio devices are only allowed to access the spectrum according to their privileges (Xu et al., 2006). The framework is based on a *trusted computing* (TC) base in each CR node that enforces the policy rules for spectrum access and etiquettes defined in the *XG Policy Language* (XGPL). Atia et al. propose an enforcement structure for defending against malicious attacks (Atia et al., 2008). The goal of the work is to provide a framework so that the primary users can distinguish between the wireless environmental losses and the presence of harmful interference of the secondary users.

A popular approach for defending against unauthorized spectrum access is to deploy a spectrum monitoring system in the CR network. The spectrum monitoring system acts as a spectrum "watch guard" for detecting spectrum misuse and carries out the following functions: (i) monitoring of the spectrum usage in a specific spatial region and over a range of frequencies, (ii) identifying wireless services and the nodes providing such services. However, design of an effective spectrum monitoring system is a challenging task since natural or man-made obstacles can change the features of the radio signal, and identification of wireless services may be difficult if an attacker can successfully emulate a specific wireless service being provided in the network. To address these problems, spectrum monitoring systems can be distributed across the nodes. Information on the wireless services in an area can be transmitted to a central monitoring location, which can, then, correlate the various inputs and check the received information against other data like the known position of the wireless services in the area and their source.

## Defense Mechanisms against Attacks on the MAC and the Cognitive Engine

Attacks on the MAC layer, network layer and on the cognitive engine of a CR network are usually defended by making a robust system design. IEEE 802.22 standard provides a robust authentication and encryption scheme to mitigate attacks on the MAC layer. As a defense mechanism for the cognitive engine, Perich and McHenry propose a policy-based spectrum access control system for the Defense Advanced Research Projects Agency (DARPA) NeXt Generation (XG) communications program for mitigating the harmful interference caused by a malfunctioning device or a malicious user for a cognitive *software defined radio* (SDR) (Perich & McHenry, 2009). The authors propose two protection mechanisms for defending against attacks on the cognitive engine. In the first approach, the authors have argued that the likely effect of a threat on a CR network is to disrupt the state machine of the CR network and to bring the CR device to an incorrect (i.e. faulty) state. Formal state-space validation, as done with cryptographic network protocols, can be applied to the state machine to ensure that a "bad state" is never arrived at. In the second approach, the authors propose that the beliefs of the cognitive engine should be constantly re-evaluated and compared to *a priori* knowledge (e.g., local spectrum regulations) or rules (e.g., the relationship between transmit power, propagation, and frequency).

## Security Mechanisms for the Cognitive Control Channels

The *cognitive pilot channel* (CPC) of a CR network is responsible for distributing the cognitive control messages. The CPC is vulnerable to numerous attacks including the DoS attacks and the saturation attacks on the control channels. A popular protection mechanism against the jamming attack in a specific spectrum band of a CR network is to use *frequency hopping*. The CPC could use more than one spectrum band and "hop" around the spectrum bands to avoid a possible jamming attack. The trade-off is an increased complexity of the CR network as the CR nodes should be notified about the change in the frequency band of the CPC. If an attacker effectively monitors the CPC, it could "chase" the CPC band for every change and eventually cause continual adaptation and outage of service to the CR network.

Yue et al. present two coding schemes for recovering lost packets transmitted through parallel channels for designing an efficient anti-jamming coding technique (Yue & Wang, 2009). The two coding schemes, known as rateless coding and piecewise coding, can be adapted to CWSNs for protecting the CPC and CCC. Meucci et al. present a lightweight mechanism for achieving security in the PHY layer in a CR network using *orthogonal frequency division multiplexing* (OFDM) (Meucci et al., 2009). In the proposed scheme, the user's data symbols are mapped over the physical sub-carriers using a permutation strategy. The security in the PHY layer is achieved using a random and dynamic sub-carrier permutation which is based on a single pre-shared information.

## Security Mechanisms using Geo-location Database of Primary Users

In this approach, the CR network provider maintains a database of the positions and transmission characteristics (e.g., transmit power) of all the primary users in the network. The CR finds its own location information using a GPS and compares the data received from the spectrum sensing functionality with the known position of the primary users. Any anomaly in position information triggers an alert for a possible malicious attack. Borth et al. propose a protection technique wherein a primary user would transmit a beacon to alert secondary users to not transmit in specific spectrum bands (Borth et al., 2008). The drawback of this scheme is that the primary user devices are to be modified for beacon transmission.

Table 2 presents a summary of various attacks on CWSNs and their respective defense mechanisms.

## OPEN PROBLEMS AND FUTURE TRENDS IN SECURITY IN CWSNS

Wireless technology is rapidly proliferating into all aspects of computing and communications. There are over 8 billion wireless devices in use today (mostly cell phones and mobile computers), and this number is expected to increase to about 100 billion by the year 2025 (Steenkiste et al., 2009). However, the anticipated exponential growth of the wireless devices and applications is contingent on our ability to design radio technologies that continue to work well with increasing deployment density – in particular, radio systems must change, and change rapidly, to cope with 2-3 orders of magnitude increase in density from 1-0-100 devices/km$^2$ today to 1000-10,000 devices/km$^2$ in 2025. Given the fact that spectrum is a finite resource, this calls for disruptive technology innovation in the radio field. Cognitive radios in general and cognitive wireless sensor networks in particular offer the promise of bringing just this disruptive technology innovation that will enable the future wireless world. Although CWSN technology have already emerged from the early stage of laboratory trials and vertical applications supports to become a general-purpose programmable radio, there is still a big gap between having a flexible cognitive radio, effectively a building block, and the large-scale deployment of CWSNs that dynamically optimize spectrum use. Building and deploying a network of cognitive radios is a complex task. The research community working on cognitive radio networks need to understand a wide range of issues including smart antenna technology, spectrum sensing and measurements, radio signal processing, hardware architectures including software-defined radio (SDR), medium access control (MAC), network discovery and self-organization, routing, adaptive control of mechanisms, policy definitions and monitoring, and learning mechanisms. Since the core focus of this chapter is on the security and privacy issues in CWSNs, we identify some security and privacy challenges that need to be addressed for large-scale adoption of these networks in real-world deployments. Following question should be addressed in this regard:

- What types of denial-of-service and other security attacks are possible using the emerging cognitive radio technology?
- Software weaknesses are known to be a major security problem in the Internet today – what are the implications of increasingly software-based radio implementations?

Table 2: Various security vulnerabilities in CWSNs and their corresponding defense mechanisms

| Attack Category | Specific Attack Type | Security Mechanism |
|---|---|---|
| Attacks on Communication Protocols | Replay attack | Use of robust authentication scheme. |
| | DoS attack | Use of frequency hopping in the cognitive control channel, code spreading etc. |
| | Malicious alteration of cognitive messages | Protection techniques based on trust or reputation, identification of masquerading threats through signal analysis, authentication of the CR nodes. |
| | Sybil attack | Use of (i) direct validation techniques of nodes including radio resource test, (ii) random key pre-distribution with identity of each node associated with the assigned keys. |
| | Hidden node problem | Data fusion process of collaborative spectrum sensing. |
| | Saturation of cognitive control channel | Robust system design and use of security scheme for protection of system integrity. |
| | Eavesdropping of cognitive radio messages | Protection of message confidentiality. |
| | Disruption of MAC, network and cognitive engine | Verification of the identities, controlled access to the resources, protection of the system integrity. |
| Masquerading Attacks | Primary user emulation attack | Protection techniques based on trust or reputation, identification of masquerading threats through signal analysis, authentication of the CR nodes. |
| | Masquerading of a secondary CR node | |
| Unauthorized Access to Spectrum | Unauthorized use of spectrum band for selfish use by an attacker | A robust and secure framework to enforce spectrum policies. |
| | Unauthorized use of spectrum band for DoS attack on primary users | A robust and secure framework to enforce spectrum policies. |
| Physical Attacks on Sensor Nodes | Physical compromise of the nodes and extraction of cryptographic credentials from the sensor nodes. | Use of (i) tamper-proof sensor hardware, (ii) trusted computing platforms in the sensors. |
| Internal Failure of Sensor Nodes | Physical failure of the sensor node hardware or Byzantine failure of sensor nodes. | Use of collaborative sensing techniques and secure and distributed data fusion algorithms. |
| Power Exhaustion Attacks on Sensor Nodes | Sleep deprivation, Frequent channel change request to drain energy | Use of robust authentication and distributed collaborative sensing. |
| Attacks on the Objective Functions on Sensor Nodes | Belief-manipulation on sensors nodes so that the overall goal optimization module of the network produces suboptimal results. | Use of techniques to enhance robustness in the sensory inputs, resilient and collaborative learning algorithms in the learning phase of the cognitive engine. |

| Attacks on Administrative Policies of Sensor Nodes | Malicious alteration in administrative policies of sensor nodes. | Protection techniques based on trust or reputation, identification of masquerading threats through signal analysis and authentication of the CR nodes. |
|---|---|---|
| | Excuse attack, Newbie-picking attack. | Use of robust authentication and trusted hardware in sensor nodes |
| Attacks on Cryptographic Protocols | Malicious attacks on the security and key management protocols. | Use of highly secure protocols with robust key distribution and management schemes. |
| Attacks on Privacy of Sensor Data | Traffic analysis, eavesdropping on sensitive sensor (i.e. source location information) | Use of (i) anonymity mechanisms, (ii) flooding – probabilistic and phantom, and (iii) onion routing. |

- How does one assure that CRs operate as intended and designed? Is there a trusted cognitive radio architecture which can address some of these security concerns?
- What authentication mechanisms are needed to support cooperative cognitive networks? Are reputation-based schemes useful supplements to conventional PKI authentication protocols?
- How the current protection techniques for spectrum management and spectrum sharing functions can be further improved? What link protection techniques could be further incorporated in the current security frameworks?
- How the performance efficiency of the different protection techniques used in collaborative spectrum sensing can be evaluated in real-world deployment scenario?
- How to design and standardize tamper-resistant module to enforce spectrum regulation policies in CR nodes and SDR devices?

When viewed from another perspective, cognitive radios offer important new capabilities to defend against intrusions or denial-of-service attacks. The spectrum sensing and SDR capability of the radio make it feasible to employ recent developments in wireless security in which physical layer properties (such as RF signatures) are used for authentication or secure communication. Also, spectrum scanning and agility associated with cognitive radios enable networks to move away from frequency channels experiencing denial-of-service attack. Location is another important feature of a wireless network, and information on geographic position can also be used to defend against certain types of attacks on cognitive networks. Some of the research issues which need to be addressed in this regard are: (i) identification of physical layer security enhancements for wireless networks and evaluation of performance in real-world deployment scenarios, (ii) evaluation of DoS attacks and methods of defense, (iii) use of geo-location for improved wireless network security, and (iv) cooperative methods for detecting and isolating intruders. While the ongoing research works on these issues are quite promising, evaluations have been mostly limited to lab environments, and it is not clear to what degree these techniques will be feasible in real-world deployments, or whether these algorithms, architectures and protocols will scale to high density environments. Large scale testing in the CR networks is mandatory for this purpose. Since the CWSNs are still in their pre-deployment phase, there is still an opportunity and a critical requirement to make security as an integral component of CR network architecture. This will require realistic practical evaluation of new techniques as they are designed and developed.

## CONCLUSION

The cognitive radio paradigm introduces entirely new types of security threats to wireless networks in general and wireless sensor networks in particular. However, wireless security in cognitive radio networks is an area that has received relatively less attention, even though security will play a key role in the long-term commercial viability of the technology. This chapter has first introduced various security threats which are common to both traditional WSNs and the CWSNs. It has then identified various additional security threats which are specific to the CWSNs and has discusses several challenges in defending against these threats. The chapter has also discussed the current security mechanisms to defend against these threats and attacks. A comprehensive taxonomy of the attacks and their respective security schemes are also presented. Some key research challenges in CR networks particularly from the perspectives of security and privacy are identified and discussed briefly at the end.

## KEY TERMS & DEFINITIONS

Wireless sensor network: A WSN consists of hundreds or even thousands of small devices each with sensing, processing, and communication capabilities to monitor a real-world environment.

Cognitive wireless sensor network: A CWSN is a WSN in which the sensor nodes have the capabilities of changing their transmission and reception parameters according to the radio environment under which they operate in order to achieve reliable and efficient communication and optimum utilization of the network resources.

Authentication: The mechanism of verification of the authenticity of a user before allowing access to the network services is called authentication.

Primary user: In cognitive radio networks using dynamic spectrum access algorithms, the primary users are those who own licenses to different frequency bands and can use those bands whenever they wish.

Secondary user: In a cognitive radio network, when the primary users are idle, another class of user called the "secondary users" can opportunistically use the spectrum on those bands. Such secondary users need spectrum sensing algorithms to detect when the primary user is active or not active.

Denial of Service (DoS): It is a class of attack in which an attacker makes the resources in the network unavailable to its legitimate users. There are different ways in which a DoS attack may be launched: jamming attack, collision attack, routing disruption attack etc.

Dynamic spectrum access (DSA): DSA is an innovative spectrum management approach in which allocation of spectrum bands for communication services can change with time or space.

Primary user emulation (PUE) attack: In this attack, an attacker creates a waveform that is sufficiently similar to that of the primary user and attacks the DSA algorithm in a cognitive radio network so that the secondary users in the communication range of the primary user erroneously conclude that the primary user is active and vacate the channel.

SSDF attack: The spectrum sensing data falsification (SSDF) attack is carried out by malicious secondary nodes in cognitive radio network to transmit false spectrum sensing data to other nodes. This attack can be catastrophic since sending of false spectrum sensing information to a data collector in the network can cause the data collector to make a wrong spectrum sensing decision.

Hidden node problem: The *hidden node problem* arises when a node of a cognitive radio network is in the protection region of a primary user but it fails to detect the existence of the primary user due to a physical obstacle or for other reasons, and transmits in the same frequency bands of the primary user, causing harmful interference.

## REVIEW QUESTIONS

1. What is a wormhole attack? Why this attack is difficult to detect in a WSN?

2. What are the different ways in which spectrum sensing can be done?

3. Explain why trust and reputation-based systems are popular in designing security systems in WSNs and CWSNs?

4. How data aggregation can be done in WSNs in a privacy-preserving manner? What is homomorphic encryption? What do you mean by fully homomorphic functions?

5. Explain three mechanisms for identification and defending against the primary user emulation attack in CWSNs.

6. What do you mean by "belief-manipulation attack" on cognitive radio networks? Why these attacks are critical in CWSNs? Discuss a few mechanisms for detecting and preventing these attacks in cognitive radio networks?

7. What types of DoS attacks can be launched on CWSNs? Discuss various mechanisms for preventing jamming attacks on CWSNs and cognitive sensor nodes.

8. What do you mean by tamper-resistant hardware? How tamper-resistant hardwares and trusted platforms can be used to defending against physical attacks on sensor nodes in a CWSN?

9. Discuss some of the well-known cryptographic protocols and key management mechanisms that can be efficiently deployed in WSNs and CWSNs?

10. Identify some of the future research challenges in the field of CWSNs in general, and security and privacy aspects of these networks in particular.

## ADDITIONAL READING SECTION